\documentclass[sigconf]{acmart}
\usepackage{alltt}
\usepackage{epsfig}
\usepackage{verbatim}
\usepackage{fancybox}
\usepackage{multirow}
\usepackage{colortbl}
\usepackage{rotating}
\usepackage{algorithm}
\usepackage{algorithmic}

\usepackage{color}
\usepackage[ampersand]{easylist}
\usepackage{cancel}
\usepackage{amsmath}
\usepackage{pgfplots}
\pgfplotsset{compat=newest}
\usepackage{tikz}
\usepackage{xcolor}
\usepackage[textsize=scriptsize]{todonotes}

\usepackage{gensymb}

\usepackage{hyperref}

\usepackage{booktabs}

\graphicspath{{./figures/}}

\newenvironment{smitemize}%
  {\begin{list}{$\bullet$}{%
      \setlength{\parsep}{0pt}%
      \setlength{\topsep}{0pt}%
      \setlength{\itemsep}{4pt}}}%
  {\end{list}}

\copyrightyear{2018} 
\acmYear{2018} 
\setcopyright{acmcopyright}
\acmConference[MobiSys '18]{The 16th Annual International Conference on Mobile Systems, Applications, and Services}{June 10--15, 2018}{Munich, Germany}
\acmBooktitle{MobiSys '18: The 16th Annual International Conference on Mobile Systems, Applications, and Services, June 10--15, 2018, Munich, Germany}
\acmPrice{15.00}
\acmDOI{10.1145/3210240.3210331}
\acmISBN{978-1-4503-5720-3/18/06}

\newcommand{\wonjung}[1]{\todo[color=yellow, inline]{\textbf{wonjung: } #1}}
\newcommand{\ken}[1]{\todo[color=green, inline]{\textbf{Kenny: } #1}}
\newcommand{\yk}[1]{\todo[color=cyan, inline]{\textbf{Youngki: } #1}}
\newcommand{\raj}[1]{\todo[color=magenta, inline]{\textbf{Rajesh: } #1}}
\newcommand{\am}[1]{\todo[color=orange, inline]{\textbf{Archan: } #1}}
\newcommand{\response}[1]{\todo[color=pink, inline]{\textbf{TODO: } #1}}
\newcommand{\missing}[1]{\textcolor{red}{#1}}

 \newcommand{\name}{{\em Empath-D\ }}
\newcommand{\names}{{\em Empath-D}}
 \newcommand{\namef}{{\em Empath-D's\ }}
 \newcommand{\namet}{Empath-D~}  
 \newcommand{\nameb}{EMPATH-D~}  

\begin{document}

\title{Empath-D: VR-based Empathetic App Design for Accessibility}

\author{Wonjung Kim}
\authornote{This work was done while the author was on an internship at Singapore Management University}
\email{wjkim@nclab.kaist.ac.kr}
\affiliation{\institution{KAIST}}

\author{Kenny Tsu Wei Choo}
\email{kenny.choo.2012@smu.edu.sg}
\affiliation{\institution{Singapore Management University}}

\author{Youngki Lee}
\email{youngkilee@smu.edu.sg}
\affiliation{\institution{Singapore Management University}}

\author{Archan Misra}
\email{archanm@smu.edu.sg}
\affiliation{\institution{Singapore Management University}}

\author{Rajesh Krishna Balan}
\email{rajesh@smu.edu.sg}
\affiliation{\institution{Singapore Management University}}


\renewcommand{\shortauthors}{Wonjung Kim et al.}

\begin{abstract}
With app-based interaction increasingly permeating all aspects of daily living, it is essential to ensure that apps are designed to be \emph{inclusive} and are usable by a wider audience such as the elderly, with various impairments (e.g., visual, audio and motor). We propose \names, a system that fosters empathetic design, by allowing app designers, \emph{in-situ}, to rapidly evaluate the usability of their apps, from the perspective of impaired users. To provide a truly authentic experience, \name carefully orchestrates the interaction between a smartphone and a VR device, allowing the user to experience simulated impairments in a virtual world while interacting naturally with the app, using a real smartphone. By carefully orchestrating the VR-smartphone interaction, \name tackles challenges such as preserving low-latency app interaction, accurate visualization of hand movement and low-overhead perturbation of I/O streams. Experimental results show that user interaction with \name is comparable (both in accuracy and user perception) to real-world app usage, and that it can simulate impairment effects as effectively as a custom hardware simulator. 
\end{abstract}
%
%



\maketitle
\section{Introduction}

Digital interactions have become increasingly commonplace and immersive. We now constantly interact with our personal devices and computing-enhanced ambient objects (such as coffeemakers, home automation systems and digital directories), while engaging in everyday activities, such as commuting, shopping or exercising. Given the ubiquity of such interactions, it is important to ensure that the associated computing interfaces remain accessible to segments of the population, such as the \emph{elderly}, who suffer from various impairments. The global elderly population is projected to reach 16.7\% by 2050~\cite{survey_elderly}, and such users suffer disproportionately from impairments (e.g., vision) that hinder accessibility.

To support more accessible design, our earlier work~\cite{hotmobile17-empathd} introduced the vision of \names, which uses a virtual reality (VR) device to provide mobile application/object designers with a \emph{realistic emulation} of the interaction experience that impaired users would encounter. In this work, we present the design, implementation and validation of the \name system inspired by this vision. \namef goal is to allow unimpaired application designers to \emph{step into the shoes} of impaired users and rapidly evaluate the usability of alternative prototypes. While we shall principally focus on empathetic evaluation of mobile applications (apps), \namef design is generic enough to permit emulation of other real-world interactions--e.g., how an elderly user with {\em cataracts} and {\em hearing loss} would experience a traffic-light controlled pedestrian intersection.

\begin{figure}[t]
\centering
\includegraphics[width=\columnwidth]{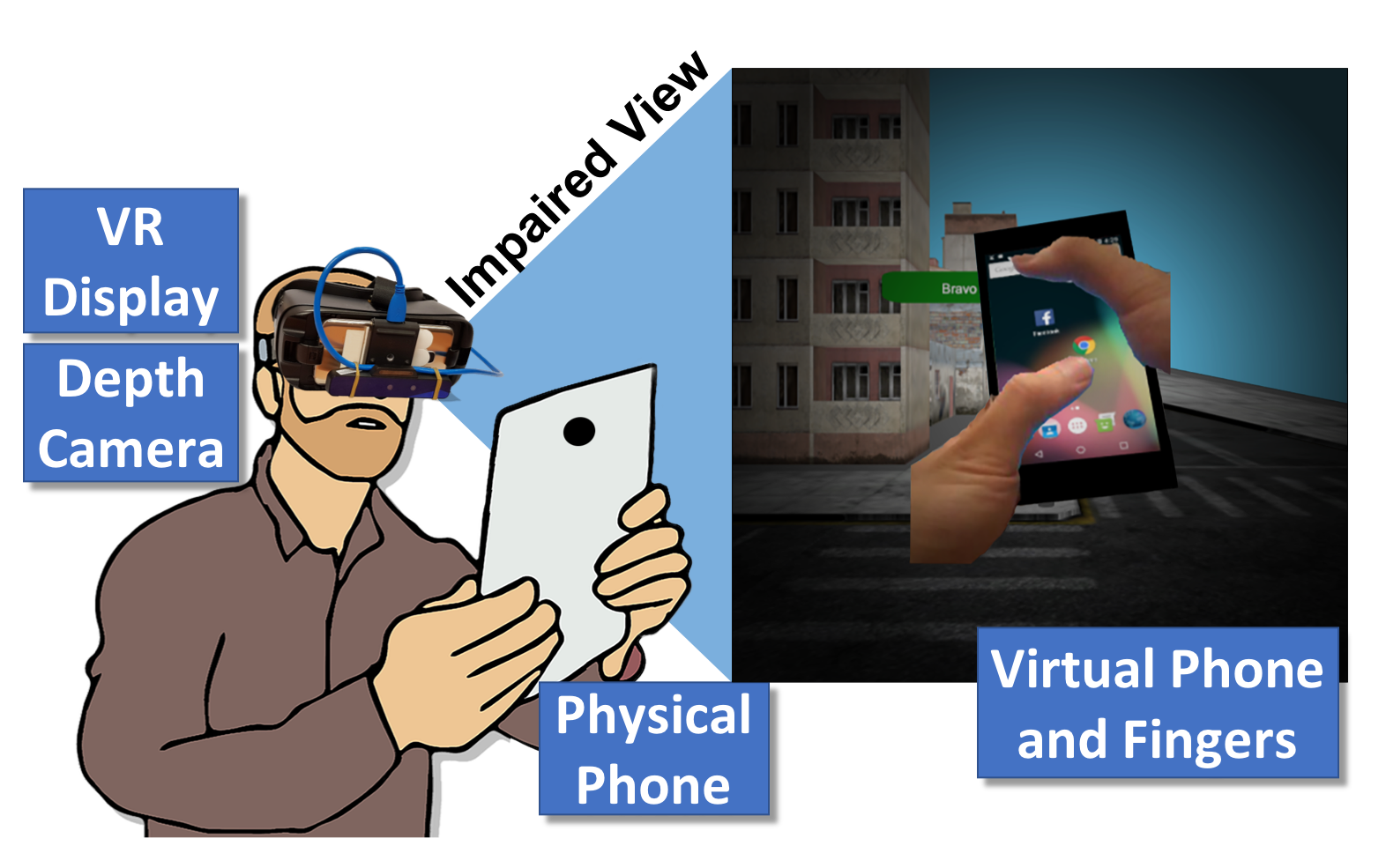}
\caption{Overview of \name }
\label{fig:empathd-overview}
\end{figure}

\namef\footnote{Video of \name in action at \url{https://is.gd/empath_d}} key idea is to present the user with an impairment-augmented view of the smartphone interface (or other digital objects) in a virtual world, while allowing the non-impaired user to perform natural interactions, using a physical smartphone, with a real-world instance of the smartphone app. At a high-level, \name works as follows (see Figure~\ref{fig:empathd-overview}): The (unimpaired) user uses a physical smartphone to perform real-world interactions (such as scrolls, taps or gestures) with the app, while wearing a VR device. The results of such interactions are projected \emph{instantaneously} through the I/O interfaces (e.g, screen, speaker) of a `virtual smartphone' visible in the VR display, but only after those I/O streams have been appropriately degraded by the specified impairment.  For example, in Figure~\ref{fig:empathd-overview}, the virtual phone's display (and the world view) has been appropriately vignetted, to mimic the experience of a user suffering from \emph{glaucoma}.

\noindent \textbf{Key Challenges:} To mimic impairments with adequate fidelity and usability, \name must support the following features:
\begin{smitemize}
\item \emph{Fast, Accurate Multi-device Operation:} \name utilizes a split-interaction paradigm: a user interacts with an app using a real-world handheld smartphone, while perceiving (viewing, hearing) the app responses through the VR interface. To faithfully replicate the real-world experience, this split-mode interaction must have tight time coupling and visual fidelity (of the virtual phone's screen), comparable to direct interactions with a standalone smartphone. 
\item \emph{Real-time Tracking:} To preserve a user's perception of naturalistic interactions, \name must not only capture explicit phone events, but also mirrors physical actions taken by the user (e.g., swinging the phone around or having one's hand hover over the phone). Thus, \name must also track and \emph{render}, in real-time, the orientation/location of both the phone and the user's hand within the VR device's field-of-view.

\item \emph{Lightweight Impairment Execution:} To preserve the feel of natural interaction, \name must insert the impairment-specific perturbations into the input/output streams with imperceptible latency or computational overhead (e.g., no reduction in video frame rate).
\end{smitemize}

\noindent \textbf{Key Contributions:}  We make the following major contributions:
\begin{smitemize}
\item \emph{3-Tier Virtualisation Model:} We design a novel 3-tier architecture where (i) the real-world smartphone serves merely as a \emph{tracker}, forwarding user interaction events (e.g., screen touch and gestures) to a computationally powerful intermediary, after which (ii) the intermediary device perturbs those events by blending in specific \emph{input} impairments (e.g., hand tremors) and passes them to an app instance running on a smartphone emulator, and finally (iii) the VR device receives the redirected outputs from this app instance and renders an appropriately-impaired (by blending in the \emph{output} impairments) virtual world, including a virtual smartphone.

\item \emph{Real-time Hand and Phone Tracking:} We use an RGB-Depth camera, mounted on the head-worn VR device, to track the outline of a user's hand, and subsequently perform a lightweight but realistic 3-D rendering of the hand on the VR display. We also use fiducial marker tracking~\cite{aruco_marker_tracking} by the camera to track the position/orientation of the real-world smartphone. We demonstrate our ability to achieve both high-fidelity (pointing error $\le 5~mm$) and low-latency (end-to-end delays below $120~msec$) hand tracking and display.
 
\item \emph{Usability of Virtualized Phone, in Use Environments:} We show that \name is not just usable, but that user performance (absent impairments) using \namef virtual smartphone is equivalent to real-world interaction with a smartphone. In addition, we allow usability testing of apps in their use environments, a key enabler for design of mobile applications which may be used anywhere. Our Samsung Gear VR-based prototype has end-to-end latency low-enough (only $96.3~msec$ of latency, excluding the mobile app emulation) to permit faithful reproduction of direct smartphone usage.

\item \emph{Validation of Impairment Fidelity and Overall System:} We implement two distinct vision (glaucoma \& cataract) and one audio (high-frequency hearing loss) impairment in our \name prototype. We then conduct a set of studies using the vision impairments, where 12 participants perform a series of standardised activities (e.g., add an alarm), using both our \name prototype (test) and a commercial hardware vision impairment simulator (control) and establish that the performance of users is \emph{equivalent} across the test and control groups. Finally, we conduct a small-scale study to provide preliminary evidence that our empathetic approach allows developers to design accessible mobile UIs faster and better.
\end{smitemize}

\section{The \nameb Vision} 
We use an example to illustrate the use of \names:

\textbf{Designing for Visual Impairment}. \textit{Alice is designing a mobile app that automatically magnifies text from real environments seen through its rear camera to aid people who suffer from cataracts (a condition that dims and blurs vision). Alice starts \name and is presented with a web interface that allows her to customise impairments (e.g., specify the intensity of visual blur). After customising the environment, Alice clicks in the \name web interface to (1) compile the environment to her phone used for VR display (VR-phone)\footnote{The VR-phone is needed only for VR devices that require a smartphone--e.g., Samsung Gear VR} and (2) connect an input/output service to a separate phone (IO-phone). She then plugs the VR-phone into the VR headset.}

\textit{Alice then compiles her Android app, and runs it in the Android emulator. She puts on the VR headset and holds the IO-phone in her hands. A virtual smartphone (Virt-phone) shows up in VR, tracking the real-world motion of the IO-phone. Alice now navigates through the virtual world, experiencing it as an ``impaired user, with cataracts".  She holds up IO-phone on a street corner (in the real world), and notices that the magnified text (as seen in the virtual phone in the virtual world) is not clear enough to be legible to a cataract-impaired user. She can now iteratively and rapidly modify her app, recompile it, and execute it in the Android emulator, until she is satisfied with the output.} This scenario demonstrates the ease-of-use for \names, with no need for special instrumentation of the app. 


\section{System Overview}
\label{sec:overview}

\subsection{Design Goals and Implications}
\label{sec:designgoals}
\name has the following key goals, which directly influence the salient implementation choices.

\begin{itemize}
    \item \textbf{Holistic emulation of impairments:} For a truly empathetic experience, the app designer must perceive the effects of impairments not just while using the mobile app, but throughout her immersion in the virtual world. Consider a user, suffering from cataract, who is interacting with her smartphone while attending a dimly dit dinner gathering. Simply blurring the phone display, while leaving the background illumination and focus unchanged, might not replicate challenges in visual contrast that an impaired user would face in reality. This requirement precludes the straightforward use of I/O redirection techniques such as Rio~\cite{amiri2014rio}, which can potentially perturb the I/O streams of only the mobile device. Instead, the impairment must be applied holistically, to the \emph{entire virtual world}.

	\item \textbf{Realistic emulation of smartphone and mobile apps in the virtual world:} \name aims at realistically emulating mobile apps within the virtual world rendered by a commodity VR headset. Realistic emulation of mobile apps imposes two requirements. (a) First, the virtual smartphone should have sufficient visual resolution, corresponding to typical usage where the smartphone is held $\approx 30cm$ away from the eye. We shall see (in Section~\ref{subsec:emurender}) that this requirement, coupled with differences in display resolutions between smartphones and VR devices, requires careful magnification of the virtual smartphone to provide \emph{legibility} without hampering \emph{usage fidelity}. (b) Second, the user should not perceive any \emph{lag} between her user input and the rendered view of the app, seen through the VR device. Quantitatively, we thus require that the task completion time, experienced by a user interacting with the emulated application in the virtual world, should be comparable to real-world app usage on a real smartphone.  

	\item \textbf{Use of unmodified app} For easy and low-overhead adoption by app designers, \name should support the emulation of mobile applications using the original, unmodified binaries (e.g., .apks for Android). \namef requirement to support empathetic emulation without app modifications implies that app designers would be able to adopt \name with minimal impact to existing development practices.

	\item \textbf{Low-latency, accurate finger tracking:} This goal is an extension of the holistic emulation objective. In the real-world, users utilise instantaneous visual feedback and proprioception to move their fingers around the smartphone display, even when they are hovering but not actually touching the display. To ensure consistency between the user's tactile, visual and proprioceptive perceptions of her hand movement, \name should also realistically render, in the virtual world, the user's hand movements and any changes in the position/orientation of the real-world smartphone, \emph{without any perceptible lag}. In Section~\ref{sec:implementation}, we shall see how the \name implementation meets these stringent performance bounds.  

	\item \textbf{Light-weight, effective emulation of impairments:} \name will need to emulate impairments, at different levels of severity. For high-fidelity empathetic emulation, the insertion of such impairments in the I/O streams of the smartphone should not add generate any additional artefacts (e.g., increased latency, reduction in display refresh rate, etc.).    
\end{itemize}

\begin{figure}[t]
\centering
\includegraphics[width=.99\columnwidth]{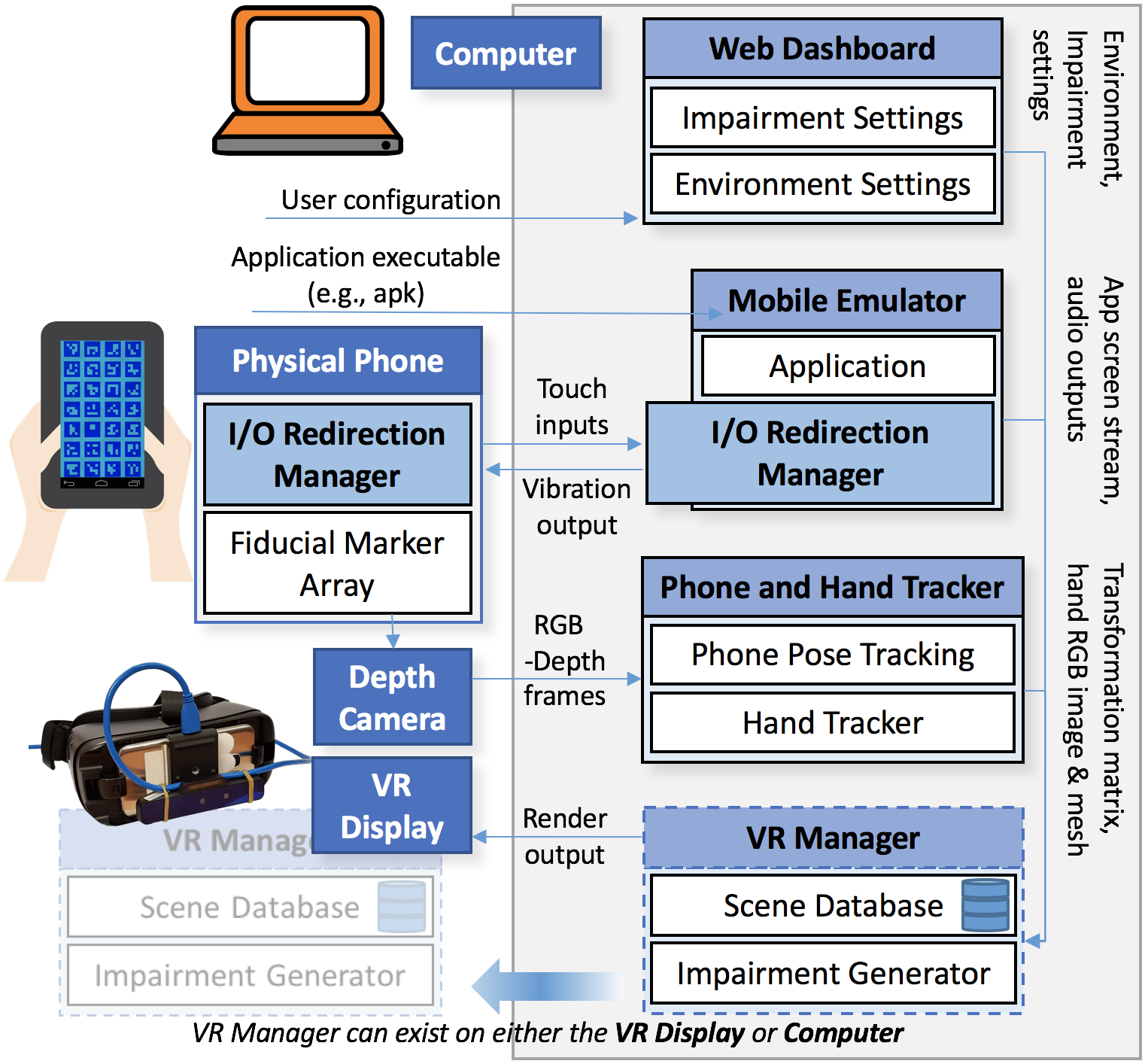}
\caption{\name architecture}
\label{fig:sys_arch_overview}
\end{figure}

\subsection{System Overview}
\label{subsec:overview}

We now present the overview of the \name system (illustrated in Figure~\ref{fig:sys_arch_overview}). 

\textbf{Using \name in VR}. To immersively evaluate the application, the developer (or the tester) starts by installing her developed application binaries (i.e., Android .apks) to run on the emulated smartphone. The developer then adjusts the profile settings for the impairment using \namef web dashboard and selects a use case scenario (e.g., in office, in the street, etc.). She holds her physical smartphone and puts on the VR headset, earphones (when hearing impairments are involved) and experiences the immersive reality (where she can use the app - now mapped onto the physical smartphone - with the configured impairment under the designated use case scenario) that \name generates. She then tests out various interfaces and functionalities of the app in the immersive VR environments.

\textbf{Components of \names}. \name runs across three different physical devices: a physical smartphone, a computer, and a VR device (see Figure~\ref{fig:sys_arch_overview}).

\noindent \emph{Smartphone:} In \names, the user interacts with the app using a real smartphone held in her hand. Interestingly, this smartphone does not run the app itself, but functions as a tracking device, helping to preserve the user's realistic sense of smartphone interaction. The smartphone simply \emph{redirects} the user interaction events (e.g., touch events such as clicks and swipes on the display and motion events captured by inertial sensors) to the computer, which is in charge of the app emulation. This smartphone also displays a fiducial marker array~\cite{aruco_marker_tracking} on its display, to help in efficient, real-time tracking of the phone's location.

\noindent \emph{Computer:} The computer is at the heart of \namef ability to fuse the real and virtual world. It consists of two major components: \emph{Phone and Hand Tracker} and \emph{Mobile Emulator}, as well as a \emph{Web Dashboard} (see Figure~\ref{fig:dashboard}), which allows the user to select the impairment profile to be applied. In addition, as we shall discuss shortly, this computer \emph{may} run an \emph{Impairment Generator} cum \emph{Virtual World Renderer}). Key functions include:
\begin{smitemize}
\item The \emph{Phone and Hand Tracker}, uses image captured by the VR headset-mounted camera, to track the position and pose of the smartphone (relative to the VR device), and create the virtual phone image at the correct position in the virtual world. It also uses the same camera to track the user's hand, as it interacts with the smartphone, and then renders it in the virtual world.

\item The Mobile Emulator executes the app being tested, using the redirected stream of user interaction events transmitted by the smartphone. The resulting visual output of the app is then transmitted as a sequence of images to the VR device, where these images will be integrated into the virtual phone object; likewise, audio output (if any) is directly streamed to the VR device.

\end{smitemize}

The overall \name framework includes an \emph{Impairment Generator} that is typically applied as one or more filters over the Virtual World Renderer (an engine such as Unity~\cite{unity}) which is responsible for combining various virtual objects and rendering the virtual world). The \emph{Impairment Generator} effectively perturbs/modifies the audio/video feeds of the virtual world, before they are displayed on the VR device.  For example, to emulate cataracts, it applies an appropriate `blurring/dimming' filter on the video feed; similarly to emulate high-frequency hearing loss (an audio impairment), this generator will apply a low-pass filter on the output audio stream. These two components are placed inside a dotted-line rectangle in Figure~\ref{fig:sys_arch_overview}, to reflect the reality that these components run on \emph{either} the Computer or the VR device, depending on whether the VR device is \emph{tethered} or not. In untethered VR devices (such as the Samsung Gear VR), the \emph{Impairment Generator} and the \emph{Virtual World Renderer} run on the VR device itself. In contrast, tethered devices such as the HTC Vive will run on the computer, and typically offer higher graphics quality, frame rates, faster execution.

\noindent \emph{VR Device:} Finally, the VR device is used to display the synthesised virtual world to the user. This synthesis involves the fusion of the virtual smartphone, the user's hand and the ambient virtual world, all subject to the impairment filter. 

\section{VR-based Emulation of mobile interaction}
\name follows a split-interaction paradigm: for realistic immersion, \name renders the visual and audio output of the target app in the virtual world (i.e., via VR headset's display and speakers), while allowing the user to interact naturalistically with a real-world physical phone. A major challenge in this paradigm is to enable natural, low-latency tracking and display of the real-world motion of both the phone and the user's hands, so as to ensure consistency across the user's visual, tactile and proprioceptive experience. We achieve this by performing three distinct steps: (a) smartphone tracking, (b) hand tracking, and (c) hand rendering in VR, using an RGB-Depth (RGB-D) camera mounted on the VR headset. \name first tracks the position and orientation of the physical smartphone and synchronises the position of the virtual phone to the physical smartphone (See Section~\ref{sec:tracking_smartphone}). Separately, \name also captures fingers in the real world and displays them at the correct position (relative to the virtual smartphone) in the virtual world (See Section~\ref{sec:hand_segmentation} and ~\ref{sec:hand_rendering}). 

\begin{figure}[th]
	\centering
	\includegraphics[width=0.65\linewidth,height=2.2in]{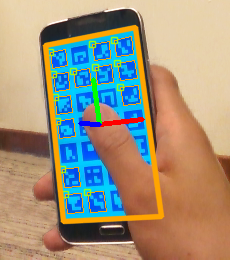}
	\caption{Tracking physical phone with fiducial markers}
	\label{fig:marker_tracking}
\end{figure}

\name uses the headset-mounted RGB-D camera to capture the colour image along with the depth values, relative to the camera. The camera's position is always fixed, relative to the user's head. Its three axes are thus aligned to a user's head: $z$-axis to the user's forward (gaze) direction, and $x$ and $y$ axes capturing the vertical and horizontal displacement. 
 


\subsection{Tracking the physical smartphone} 
\label{sec:tracking_smartphone}

\name uses fiducial markers, displayed on the physical smartphone's screen, to localise the smartphone efficiently. It takes a colour image as an input, and returns the transformation relative to the camera's coordinate system: translation and rotation, i.e., x, y, z, roll, pitch, yaw from the RGB-D camera's coordinate system. We employ a technique proposed and detailed in~\cite{aruco_marker_tracking}. 

The \name Hand Tracker component tracks the physical phone using markers captured by the camera. Each marker, displayed on the phone screen, has a distinct pattern. The tracker knows the position of each marker  (e.g., top-left, top-right, bottom-left and bottom-right) in the physical smartphone screen's coordinate system. The system first detects these markers in a given colour image, identifying them based on their unique patterns (see Figure~\ref{fig:marker_tracking}). In particular, the system recognises the coordinates of each of the four corners of each marker. Moreover, the system knows the true size of, and separation between, each marker. It then uses an object pose estimation algorithm (provided by openCV's \emph{solvePnP} function~\cite{solvepnp}), along with the array of fiducial marker points, to compute the 3-D position and orientation of the smartphone. Past results~\cite{aruco_marker_tracking} show that this technique can compute an object's position and orientation with sub-cm level accuracy.

This fiducial marker-based algorithm would fail under two conditions: (a) when the markers are occluded by the user's hand, and (b) if the ambient illumination levels are too low or too high, reducing the contrast level of the markers. To tackle (a), the smartphone screen uses an entire array of markers displayed across the scene, thereby ensuring correct smartphone tracking as long as some part of the phone is visible. Contrast concerns are not particularly relevant in our scenario, as we assume that the user is testing the app in a regularly lit work/office environment.

\subsection{Hand Segmentation}
\label{sec:hand_segmentation}

\begin{algorithm}[t]
	\caption{Hand Segmentation}
	\label{alg:segmentation}
	\begin{algorithmic}[1]
		\STATE \textbf{Input: }${T \gets }$ \textit{Phone's translation} (3-D vector)
		\STATE \textbf{Input: }${R \gets }$ \textit{Phone's orientation} (3$\times$3 rotation matrix),
		\STATE \textbf{Input: }${F \gets}$ \textit{RGBD Frame}, 2-D array that each entry $F_{i, j}$ holds a color value and 3-D position relative to the camera.
		\STATE \textbf{Input: }${V \gets}$ \textit{3-D region of interest} (relative to the phone)
		\STATE \textbf{Output:} $fgMask$, 2D bool array whose dimension equals to $F$
		\STATE 
		
		\STATE $fgMask[i, j] \gets \FALSE \text{ for all } (i, j)$
		\FOR {\textbf{point} $(i, j)$ \textbf{in} $F$}
			\IF {$(i, j)$ \text{in} \textit{screen\_border}}  
				\STATE /* Case A: Blue background segmentation */
				\STATE {$fgMask[i, j] \gets 1 - Blue(F_{i, j}) + 0.5 \cdot Red(F_{i, j}) > \tau$} \label{alg:line:color_segmentation}
			\ELSE
				\STATE /* Case B: Depth-based segmentation */
				\STATE {$pos_{phone} \gets R^{-1} \cdot (Position(F_{i, j}) - T)$}
				\STATE ${fgMask[i, j] \gets (pos_{phone} \in V)}$
			
			\ENDIF
		\ENDFOR
		\STATE {$\text{\textbf{return }} fgMask$}	
	\end{algorithmic}
\end{algorithm}

\name uses the frames captured by the RGB-D camera to track and segment the user's hand. For each frame, we extract the segment (polygon of pixels) that represents the user's hand, and render that segment in the virtual world. As the goal of hand-tracking is to provide the user with a natural view of her smartphone interactions, we restrict the tracking technique to a 3-D region of interest (ROI) that is centred at the phone, with a depth of $2cm$ and a planar boundary of $6cm$. In other words, we only track the hand while it is $\le 2cms$ away from the smartphone screen, and within $\le 6cms$ of the smartphone edges.

%

\begin{figure}[th]
	\centering
	\begin{minipage}{0.45\columnwidth}
	\centering
	\includegraphics[width=0.9\columnwidth, height=1.8in]{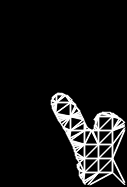}
	\caption{Mesh of hand\\~}
	\label{fig:hands_mesh}
	\end{minipage}
	\begin{minipage}{0.45\columnwidth}
		\centering
		\includegraphics[width=0.95\columnwidth,height=1.8in]{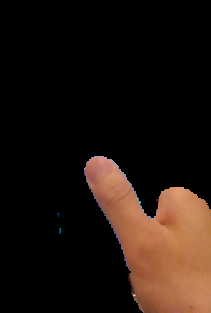}
		\caption{\name hand segmentation}
		\label{fig:segmented_hands}
	\end{minipage}
\end{figure}

A straightforward approach is to apply a depth-based segmentation strategy, where we first isolate only the foreground points which lie within a depth=$2cm$ of the smartphone surface. However, we empirically observed that, due to the glossy surface of the smartphone, such depth estimation was inaccurate for points located on the smartphone's screen. Accordingly, we implemented two separate segmentation methods (detailed in Algorithm~\ref{alg:segmentation}): (case A) a colour-based segmentation approach to identify points which are directly over the smartphone, and (case B) a depth-based approach to identify points which are near, but not over, the smartphone's screen. We apply the colour-based segmentation to the points inside the screen's border (thick orange contour in Figure~\ref{fig:marker_tracking}) and the depth-based approach to the points outside.

\noindent \emph{Colour-based segmentation:} 
We adopt the colour-based technique proposed in~\cite{blue_screen_matting}.
The approach tests RGB values to segment foreground (hand) from background, coloured in blue. 
In our scenario, we target human skin as the foreground. Human skin has a property common in all races: its R value has about twice the value of G and B ($R \approx 2G \approx 2B$).
Given the property of human skin, we obtain a formula that discriminates the foreground from the background whose $B$ value is 1 (line~\ref{alg:line:color_segmentation} in Algorithm~\ref{alg:segmentation}). $\tau$ is a user-tunable threshold which allows it adapt to different lighting conditions.

However, note that, to enable tracking of the phone, the phone's screen cannot be completely blue, but will need to contain the array of fiducial markers.
We tackle both problems simultaneously by using blue ($R$=0, $G$=0, $B$=1) to colour the markers, over a cyan ($R$=0, $G$=1, $B$=1) background. Here we modified only $G$ value, which is unused in the colour-based segmentation.



Points outside the smartphone's screen are segmented using the depth-based approach. After identifying the points corresponding to the user's hand, the system translates these points to 3-D coordinates in the camera's coordinate system, using
the associated depth values. 



\subsection{Rendering the hand in the virtual world}
\label{sec:hand_rendering}
After detecting the hand segment, the \name system renders it in the virtual world. The system passes the tracked hands to the \emph{Virtual World Renderer}, sharing the (i) 3D structure of the hands (surface mesh), (ii) colour image of the RGB-D frame (texture), and (iii) mapping between the surface mesh and the colour image (UV map). In common rendering engines (e.g. Unity), the 3D structure of the hand is represented by a \emph{triangle mesh}--i.e., a set of vertices, constituting individual small triangles. The mesh is rendered at the same location as the user's hand in the real world. As the user's hand is localised in the coordinates of the RGB-D depth camera, the location is offset by an additional depth value (7cm in our implementation), to reflect the additional distance between the centre of the user's eyes and the depth camera. An important characteristic of our algorithm is that we render the \emph{actual image} of the user's hands over this triangle mesh. 
Figure~\ref{fig:hands_mesh} illustrates the Delaunay triangulation of a set of points. The mesh is combined with the hand's image (Figure~\ref{fig:segmented_hands}), and rendered in the VR display. Extracting and rendering the actual image of the user's finger enhances the immersive feeling of real-life smartphone navigation in the virtual world.

The complexity of the mesh--i.e., the number of vertices (or triangles) in the rendered hand--is an important parameter in the rendering process. A larger number of vertices captures the contours of the hand more precisely, resulting in a more life-like image. However, this also results in added rendering latency in the rendering engine. To support the twin objectives of low-latency and life-like rendering, we utilise a sub-sampling technique to construct the mesh. Specifically, \name preserves all the points on the edges of the segment, to preserve the precise contours of the hand. However, it performs a 32-fold downsampling of the interior points (prior to constructing the Delaunay triangulation), along both the row and column axes, to reduce the computational time significantly, without materially affecting the reconstructed hand image. We shall show, in Section~\ref{sec:implementation}, how our prototype \name implementation uses this technique to achieve our twin objectives.

\section{Impairment simulation}
\begin{figure}[t]
	\centering
	\includegraphics[width=0.9\columnwidth]{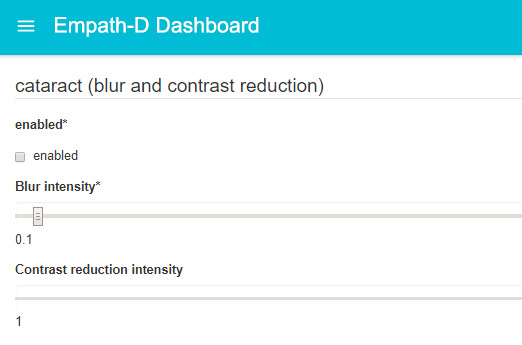}
	\caption{Screenshot of \name impairment configuration dashboard }
	\label{fig:dashboard}
\end{figure}

\name aims to enable evaluation of the usability of app designs under visual, auditory and haptic impairment simulation. Realistic simulation of various impairments in the VR world is the essential requirement to achieve this goal.

There has been a thread of research to simulate impairments through physical simulator devices~\cite{rousek_simulating_2009, zimmermanlowvision, lowvisionsimulator, agnes, gert}. For instance, Zimmerman et al. use goggles and enclosing materials to simulate low vision impairments~\cite{zimmermanlowvision}. These hardware simulators generalise the impairment of interest and enable simulation of specific aspects of the impairment pathology rather than emulate exactly how an impairment is. However, impairments can vary greatly between individuals. For instance, glaucoma generally progresses in deterioration from the periphery towards the centre of vision, but in reality, it comes in different shapes and severity, affecting usability of applications in different ways. Existing physical impairment simulators simply approximate this as a central circle of clarity, with blur through to the periphery. \name is advantageous over existing physical simulators in the following ways, it allows: 1) impairments to be customised, 2) simultaneous manifestation of multiple impairments, 3) the addition of new impairments easily. Figure~\ref{fig:dashboard} shows the web interface for designers to customise impairments for the target user group.

\subsection{Simulating Visual Impairments}

\begin{figure}[t]
	\centering
	\begin{minipage}{0.49\columnwidth}
		\centering
		\includegraphics[width=0.95\columnwidth, height=1.5in]{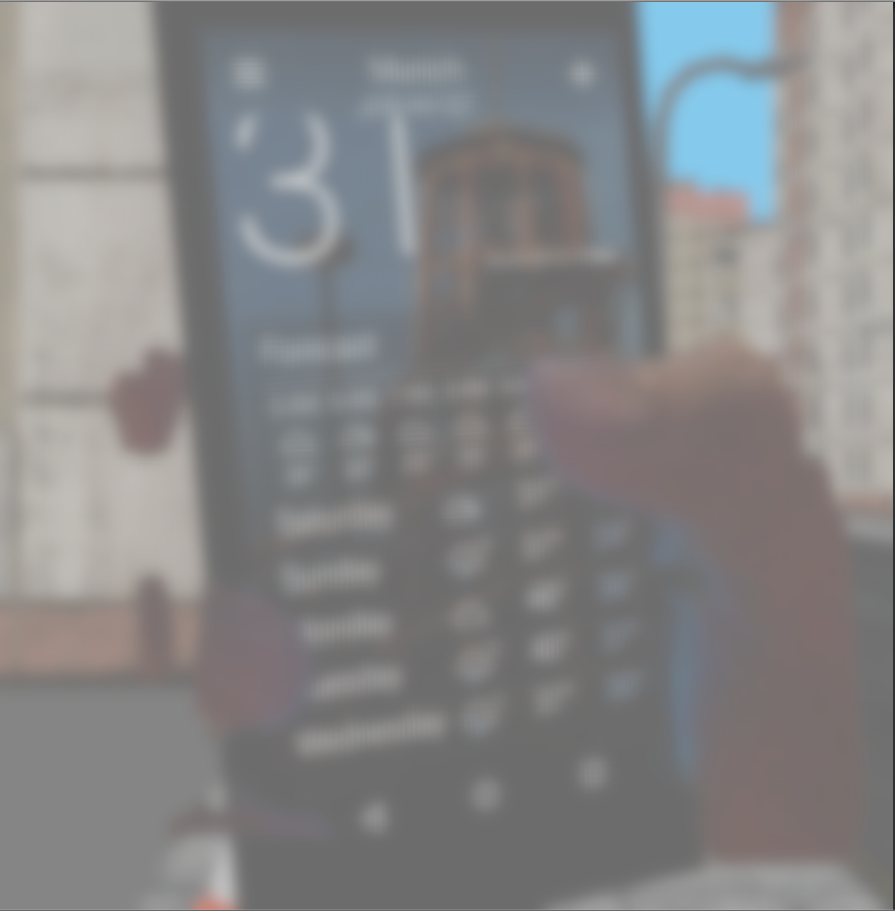}
		\label{fig:cataract}
	\end{minipage}
	\begin{minipage}{0.49\columnwidth}
		\centering
		\includegraphics[width=0.95\columnwidth,height=1.5in]{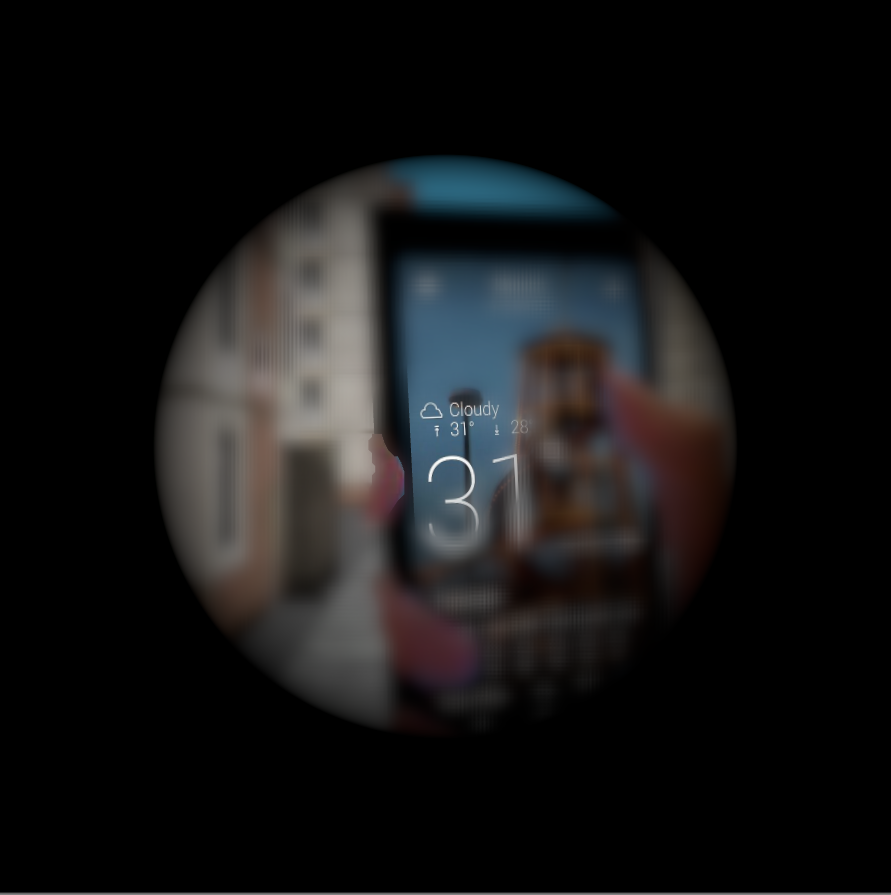}

		\label{fig:glaucoma}
	\end{minipage}
	\caption{Simulated cataract (left) and simulated glaucoma (right)}
	\label{fig:impairment_simulation}
\end{figure}

Vision is the dominant sensory system by which humans perceive the world, and is a key focus for \names. Vision impairment is one of the most common causes of accessibility problems that comes with age. Common vision impairments include cataracts, glaucoma, and age-related macular degeneration. Such vision impairments present as reduced visual acuity, loss of central/peripheral vision, or decreased contrast sensitivity. It is widely studied that these symptoms can affect the interaction with various desktop and mobile applications; for example, humans use peripheral vision to pre-scan text ahead of his/her point of focus. As the peripheral vision narrows, the scanning becomes less effective, which slows reading~\cite{legge2007case}. In this work, we examine and simulate two commonly found visual impairments - cataracts and glaucoma.

Our approach is to apply an image effect at the ``eye'' (i.e., a camera pair of view renderers) of the VR scene. From this camera pair, the image effect will apply to all other objects in the scene (e.g., smartphone, fingers, scene), just as how impaired users would experience it. We employed various image filters for different impairments, which  1) provide realism of impairments to help designers to find out usability issues and take corrective actions, and 2) have small computational overhead not to add noticeable delays to our entire emulation.

The approach is flexible and lightweight. Impairment simulator's intensity is configurable at runtime. The image effects are applied at the last stage of the rendering pipeline.
\textbf{\textit{Glaucoma}} presents functionally as a \textit{loss in peripheral vision}. To simulate glaucoma, we use a vignette with a clear inner circle, blurred inner-outer circle, and black extending outwards from the outer circle (see Figure~\ref{fig:impairment_simulation}). \textbf{\textit{Cataracts}} presents functionally as \textit{reduced visual acuity} and \textit{reduced contrast sensitivity}. We use a blur filter to simulate reduced visual acuity, and a contrast reduction filter to simulate reduced contrast sensitivity (see Figure~\ref{fig:impairment_simulation}).

The functional aspects of vision impairments are straightforward to create in VR, which give \name high extendability to implement other types of visual impairments. While we just described two impairments pertaining to our studies, it is easy to create other impairments such as colour filters to simulate colour blindness. However, we leave the effect of eye movements on impairments as the future work. Since eye-tracking is currently not supported in \names, a user will need to move his head to achieve the same effect.

\subsection{Simulating Other Modalities}
We discuss how other modalities may be simulated in \names.

\textbf{Hand Tremors}. Hand tremors are a common symptom of Parkinson's disease or Essential tremor and make it hard for one to precisely point on a touchscreen. A hand tremor may be characterised by the frequency and amplitude of oscillatory movement. Since we present virtual representations of the user's hand (i.e., as a 3D mesh) to enable his interaction with the virtual mobile phone, \name similarly perturbs this 3D mesh in VR to create hand tremors. While a user may physically not experience hand movement, the visual perturbation would be sufficient to hinder accurate touch to simulate hand tremors.

\textbf{Hearing Loss}. High-frequency hearing loss is a common symptom for the elderly population. People diagnosed with high-frequency hearing loss are unable to hear sounds between 2,000 Hz and 8,000 Hz. These people often struggle to understand or keep up with daily conversations (missing consonants in higher registers, such as the letters F and S or female voices). \name applies a bandpass filter over the output sound of the target application to diminish the sound signals between 2kHz and 8kHz and plays the filtered audio feed through the VR device.

\section{Implementation}
\label{sec:implementation}
\begin{table}
	\centering
	\caption{Hardware of \namet}
	\begin{tabular}{|c|l|}\hline
		VR headset & Samsung Gear VR~\cite{gearvrspecs}\\\hline
		VR smartphone & Samsung Galaxy S7~\cite{galaxys7specs}\\\hline
		RGB-D camera & Intel RealSense SR300~\cite{intelsr300specs}\\\hline 
		   & CPU: 4 cores, 3.4 GHz\\
		PC & RAM: 16 GB\\
		   & GPU: GeForce GTX 1080~\cite{geforce1080specs}\\\hline

		Physical IO smartphone & Samsung Galaxy S5~\cite{galaxys5specs}\\

		 \hline
	\end{tabular}
	\label{tab:system_hardwares}
\end{table}


\subsection{Hardware}

We implemented our current \name prototype using the hardware described in Table \ref{tab:system_hardwares}. We used the Samsung Gear VR fitted with the Samsung Galaxy S7 as the VR headset. We used the Intel RealSense SR300 RGB-D camera for finger tracking, selecting this among alternatives as: 1) its small size and low weight allowed us to easily attach it to the VR headset, and 2) its minimum sensing range is low enough to permit hand tracking at a distance of 30cm. We employed the Samsung Galaxy S5 as the physical I/O device, and a powerful laptop (4 core 3.4 GHz CPU, 16GB RAM) as the intermediary device. The choice of the VR headset itself was deliberate. We chose a Samsung Gear VR headset (an untethered smartphone-powered VR device) over more powerful PC-tethered VR devices such as the HTC Vive or Oculus Rift. This was mainly because PC-tethered devices such as HTC Vive use IR lasers to localise the headset, which interferes with the IR laser emitted by the RGB-D camera used for depth sensing in  hand tracking.


\subsection{Rendering an Emulated App}
\begin{figure}[!t]
	\centering

	\begin{tikzpicture}

			\begin{axis}[
			xlabel={Virtual display width (px)},
			ylabel={Rendering framerate (FPS)},
			ymin=0, ymax=60,
			height=0.5\columnwidth,
			ymajorgrids=true,
			grid style=dashed,
			]
			
			\addplot[dashed, 
				color=black,
				mark=*,
				mark size=1.5pt
			]
			table [x={width}, y={FPS}] {data/vdisfps.dat};
		
		\end{axis}
	\end{tikzpicture}
	\caption{Rendering frame rate under varying virtual display resolution (width : height = 9 : 16, default resolution of Android emulator is 1080x1920)}
	\label{fig:virtualdisplayfps}
\end{figure}
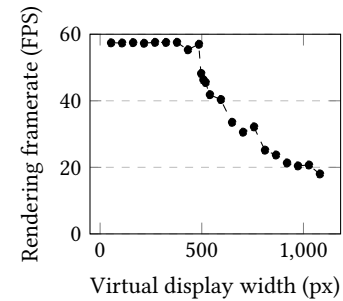

We used empirical studies to determine an appropriate screen resolution and frame rate to render the emulated app (and the smartphone) in the VR headset. \name obtains screenshots of its mobile emulator using the Android virtual display~\cite{androidvirtualdisplay} and transmits these screenshots over WiFi to the Gear VR device. The overhead of transmitting and rendering these emulated screenshots is proportional to their resolution. The default 1080p resolution could sustain a frame rate of only 18 $fps$, which causes visible jerkiness.  To reduce this overhead, we reduced the resolution (using \emph{setDisplayProjection()} method), and applied \emph{differential transmissions}, sending a screenshot only when the emulated app's display changes. 

Figure~\ref{fig:virtualdisplayfps} shows the experimental results on the tradeoff between the resolution and the rendering frame rate, obtained while playing a video to ensure continuous change of the screen content. The frame rate saturates at 57 $fps$, at a screen resolution of 485$\times$863. Moreover, through another user study (described next) to understand the minimum resolution to read an app's contents, we empirically verified that the participants had no issues in reading the app's content at the resolution of 485$\times$863. Hence, we choose this resolution as our default, although this setting can be modified (e.g., we can pick a higher resolution, and a lower frame rate, for an app with mostly static content).

\begin{figure}[t]
	\centering
	
	\begin{tikzpicture}

	\begin{axis}[
	xlabel={magnification ratio},
	ylabel={font size (sp)},
	ymin=5, ymax=20,
	height=0.5\columnwidth,
	width=\columnwidth,
	ymajorgrids=true,
	grid style=dashed,
	error bars/error bar style={
		thick,
	},
	error bars/y dir=both,
	error bars/y explicit,
	]	
	
	\addplot[dashed, 
	color=black!90,
	mark=*
	]
	table [x={factor}, y={avg_min_font_size}, y error={stddev_min_font_size}] {data/magnification_font_size.dat};
		
	\addplot[densely dotted, 
	color=black,
	mark=triangle,
	]
	table [x={factor}, y={avg_clear_font_size}, y error={stddev_clear_font_size}] {data/magnification_font_size.dat};
	\legend{min. readable font size, clearly readable font size}
	\end{axis}
	
	\end{tikzpicture}
	\caption{Readable font size of the virtual smartphone at a magnification ratio}
	\label{fig:virtualphone_magnification}
\end{figure}
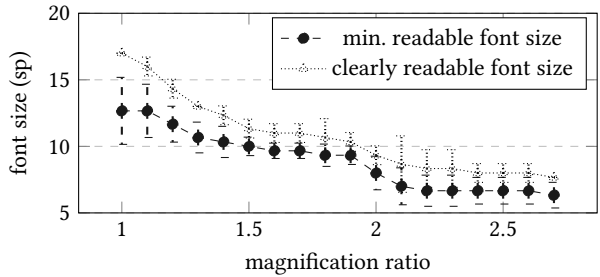
If \name displays the virtual smartphone at its original size in the virtual world (portrait position), its display becomes illegible. For example, the Samsung Galaxy S7 (in the Gear VR) has a resolution of 2560$\times$1440 and an $\approx101 \degree$ horizontal field of view yielding a horizontal pixel density of $\approx25.3$ pixels/degree.
When a virtual phone is held at 30cm away, the horizontal pixel density drops below 25.3 pixels/degree due to downsampling of the virtual phone screen as seen through the VR display. This presents a problem for viewing the content of the virtual phone --- in particular, text --- as its pixel density is significantly lower than when viewing a physical phone. For instance, the Galaxy S5 gives $\approx 89.4$ pixels/degree at 30cm distance.



We tackle this issue by scaling up the virtual phone's size by a factor that ensures that the phone's display text remains legible. To determine this factor, we recruited three participants and asked them to record the minimum readable font sizes, while showing them a virtual smartphone (at a distance of 30 cm) with various magnification ratios (increased by 0.1 from 1.0 to 2.7). Figure~\ref{fig:virtualphone_magnification} shows that participants could read text with the font size$=12$sp (the commonly used minimum font size for mobile apps) for magnification factors $\ge 1.5$. Accordingly, we used 1.5 as the default magnification ratio for the smartphone and its display. We also proportionately scaled up the user's rendered hand. User studies (Section~\ref{sec:evaluation}) show that users found this configuration highly usable.




\subsection{Rendering Virtual Hand}
\label{subsec:emurender}

\begin{figure}[t]
	\centering
	\begin{tikzpicture}
	\begin{axis}[
		xlabel={Image quality (SSIM score)},
		ylabel={Rendering latency (ms)},
		height=0.5\columnwidth,
		width=0.9\columnwidth,
		ymajorgrids=true,
		grid style=dashed,
		error bars/error bar style={
			thick,
		},
		error bars/y dir=both,
		error bars/y explicit,
		ylabel style = {align=center},
		nodes near coords, 
		point meta=explicit symbolic,
	]	
	
	\addplot [dashed, 
	color=black,
	mark=*,
	   point meta=explicit symbolic,
	   point meta/.append style={font=\footnotesize},
		nodes near coords,
	]
	table [
		x={ssimmean}, 
		y={renderinglatencymean}, 
		y error={rendering_latency_stddev}, meta=meta] {data/hand_rendering_latency.dat};
	\end{axis}

	\end{tikzpicture}
	\caption{Rendering latency vs. image quality of the virtual hand}
	\label{fig:virtualhand_rendering_latency}
\end{figure}
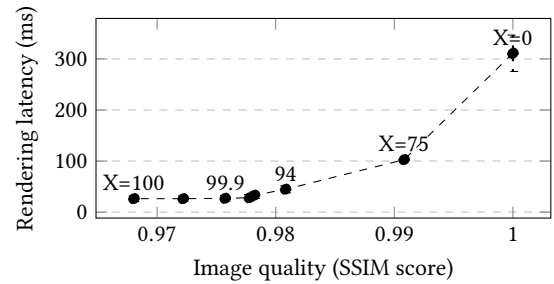

As discussed in Section~\ref{sec:hand_rendering}, the rendering latency of the virtual hand is proportional to the number of vertices in the Delaunay triangulation-based mesh. To reduce the latency, we apply a non-uniform sampling approach. Specifically, \name preserves all the points on the edges of the segment, to preserve the precise contours of the hand. However, it performs a downsampling of the interior points (prior to constructing the Delaunay triangulation), along both the $x$ and $y$ axes, to reduce the computational time significantly, without materially affecting the reconstructed hand image. We empirically determined the sampling rate $X$, by varying $X$ and measuring both (i) the processing latency and (ii) the SSIM~\cite{wang2004image, cuervo2015kahawai} (\emph{Structural SIMilarity}; a metric of perceived image quality) of the hand images, using 200 RGB-D frames. Figure~\ref{fig:virtualhand_rendering_latency} shows the results. Without any subsampling ($X=0\%$), the rendering latency is $311.1~msec$, which is too high for our responsiveness goal. We empirically downsample the internal pixels by a factor of $32$ ($X=99.9\%$), i.e., choosing every $32^{nd}$ pixel on the grid. This results in a latency of $26.9 ~msec$, while keeping the $SSIM=0.976$, a level indistinguishable with the original as perceived by a human.

\subsection{Environment Emulation}
To enable holistic evaluation of app interactions, \name emulates not just the virtual phone, but the entire virtual world as well. In our current implementation, we emulated a crowded \textit{Urban Street} environment, which includes crosswalks, traffic lights, pedestrians and commonplace roadside obstacles. To further mimic real-world movement, our implementation allows the user to navigate the virtual world by (i) rotating her head (this uses the head tracking ability of the VR device), and (ii) by `walking in place', using the technique proposed in~\cite{tregillus2016vrstep} as this does not require any additional hardware on the VR device. 

\subsection{VR Manager} 
This component currently executes on the VR smartphone, and is responsible for combining the output of the various components (\emph{Hand Tracker}, \emph{Phone Tracker} and \emph{Virtual Phone}) in the virtual world. This component, implemented as a Unity application, renders these various components.  This component is also responsible for applying the impairments on the output of the virtual world. The image effects simulating low vision impairments are defined as a script, \emph{Shaders} in Unity.

\section{Evaluation}
\label{sec:evaluation}

We now present a mix of system and user experiments to evaluate the performance and efficacy of our \name implementation. Besides micro-benchmark studies, we conducted two experiments to capture user interaction with \names. In Experiment 1, we examine the performance of \name vs. a real-world smartphone, in the absence of any impairments. In Experiment 2, we consider an impairment-augmented version of \names, comparing the performance of users against the use of commercial impairment simulation hardware. 

\subsection{Micro-benchmark Performance of \name}
We measured the overall latency of \names, both in terms of the delay in reflecting touch interactions in the virtual world and in terms of the hand tracking delay.

\subsubsection{End-to-end Latency of Touch Interaction}
As a measure of the overall responsiveness of \names, we computed the latency between a touch input, on the physical smartphone, and the resulting change in the content of the virtual smartphone, rendered in the VR display. To measure this, we utilised a high framerate camera (operating at $240~fps$) to concurrently record both the screen of the physical smartphone and the virtual phone (displayed in the VR). The phone screen is coloured green initially, and was programmed to turn red as soon as it received a touch input. We repeated the measurement 23 times, capturing (via the video frames) the time gap between (i) the physical smartphone screen turning red and (ii) the virtual smartphone turning red in the VR display. The end-to-end latency is $237.70~msec$ ($SD=20.43$). 

By monitoring the intermediary computer, we obtained the breakdown of this delay: (i) smartphone responsiveness (the time from the user touching the screen till the time the phone transmits the touch event to the computer) $=0.3~msec$ ($SD=0.16$); (ii) computer emulation responsiveness (the time from receiving the touch event till the time the screenshot of the modified display is sent to the VR device) $=141.37~msec$ ($SD=6.6$), and (iii) the VR responsiveness (the time from receiving the screenshot till it is rendered on the VR display) $=10.46~msec$ ($SD=8.36$). The remaining latency ($\approx 87~msec$) can be attributed as the WiFi network latency. These micro-measurements suggest that the default Android emulator used in our studies was the dominant component of the latency. The default Android emulator is known to be fairly slow, and multiple third party emulators (e.g., Genymotion~\cite{genymotion}) are reported to provide significantly lower latency. Accordingly, we anticipate that this overall latency can be reduced to $\le 150~msec$, without any significant architectural modification of \names.

\subsubsection{End-to-end Latency of Virtual Hand}
We also evaluated the latency between the physical movement of the user's hand and the rendering of this movement in the VR display. To capture this time difference, we displayed a small circle, at a specific point on the display, on both the smartphone and the virtual phone.  Users were instructed to swipe a finger on the screen to reach the circle. We measured, over 20 experiments, the time (no. of frames from the previously used high framerate camera) between the occlusion of the circle on the physical phone and the resulting occlusion in the virtual phone, computing an average latency of $117.46~msec$ ($SD=20.44$). Additionally, we measured the component delays of this rendering process as: (i) reading an RGBD frame: $4.90~msec$ ($SD=0.58$); (ii) phone tracking: $4.56~msec$ ($SD=0.25$); (iii) hand tracking: $8.0~msec$ ($SD=1.58$), and (iv) the VR responsiveness (the time from receiving the hand mesh till it is rendered on the VR display): $26.99~msec$ ($SD=5.22$).
The remaining latency, attributable to the WiFi network, is $\approx73$ msec, consistent with the measurements reported above.

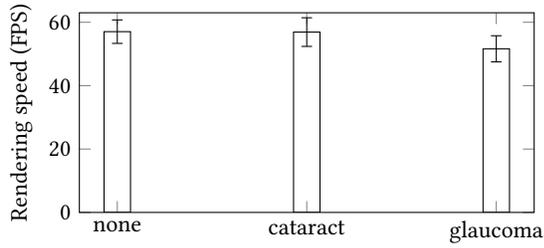
\begin{figure}[!t]
	\centering
	\begin{tikzpicture}
	\begin{axis}[
	  symbolic x coords={none, cataract, glaucoma},
	xtick=data,
	ylabel={Rendering speed (FPS)},
		ymin=0, ymax=63,
		error bars/y dir=both,
	error bars/y explicit,
	width=.9\columnwidth, height=.5\columnwidth,
		]

	


	\addplot[ybar, ] coordinates {
		(none, 57.04166667) +- (0, 3.706446425)
		(cataract, 56.90833333) +- (0, 4.520719496)
		(glaucoma, 51.63333333) +- (0, 4.119911342)
	};
	\end{axis}
	\end{tikzpicture}
	\caption{Overhead of impairment simulation}
	\label{fig:impairmentsimulationverhead}
\end{figure}

\subsection{Study Design for Usability Experiments}
We then conducted user studies on the usability and real-world fidelity of our \name implementation. The user study (approved by our institution's IRB) consisted of 12 users (9 males) with no pre-existing uncorrected vision impairments. Users were aged 24--39, with a mean age of 30.3 years (SD = 5).

\textbf{Study Tasks and Measures}. We adopted a repeated measures design, with participants counterbalanced for condition order (see Table~\ref{tab:expt1taskblock} for the conditions). Participants were asked to perform four different tasks split into two task types; \textit{everyday phone use}, and \textit{controlled pointing} (see Table~\ref{tab:expt1tasks}). Users were asked to perform all tasks using two-handed interaction, holding the phone at a distance that they normally would during daily use. We chose two-handed interaction to eliminate for phone balancing that is typical in one-handed interaction given the typical size of today's smartphones.

\begin{table}
	\centering
	\caption{Study Tasks and Conditions in Experiment 1}
	\begin{tabular}{|c|c|c|c|c|}
		\hline
		\rowcolor{gray!50}
		& \textbf{Cond} &  & \textbf{Simulator} & \textbf{Enviro}\\
		\rowcolor{gray!50}
		\multirow{-2}{*}{\textbf{Task}} & \textbf{-ition} & \multirow{-2}{*}{\textbf{Impairment}} & \textbf{Type} & \textbf{-nment}\\
		\hline
		& A & none & none & Real\\\cline{2-5}
		\multirow{3}{*}{T1~} & B & Cataracts & Physical & Real\\\cline{2-5}
		\multirow{3}{*}{~-T4} & C & none & none & Virtual\\\cline{2-5}
		& D & Cataracts & Virtual & Virtual\\\cline{2-5}
		& E & Glaucoma & Real & Physical\\\cline{2-5}
		& F & Glaucoma & Virtual & Virtual\\
		\hline
	\end{tabular}
	\label{tab:expt1taskblock}
\end{table}

\begin{table}
	\centering
	\caption{Smartphone Interaction Tasks in Experiment 1}
	\begin{tabular}{|l|c|l|}\hline
		\rowcolor{gray!50}
		\textbf{Task} & \textbf{Task}	&	\\
		\rowcolor{gray!50}
		\textbf{Type} & \textbf{Code} & \multirow{-2}{*}{\textbf{Task Description}}\\\hline
		\multirow{2}{*}{Everyday} & T1			&	Perform a Calculation\\\cline{2-3}
		\multirow{2}{*}{Phone Use} & T2			&	Add an Alarm\\\cline{2-3}
		 & T3			&	Search, Save Image on Browser\\\hline
		Controlled & \multirow{2}{*}{T4}	&	\multirow{2}{*}{Number Search and Point}\\
		Pointing & & \\\hline
	\end{tabular}
	\label{tab:expt1tasks}
\end{table}

\begin{figure}[t]
	\centering
	\includegraphics[width=0.4\columnwidth]{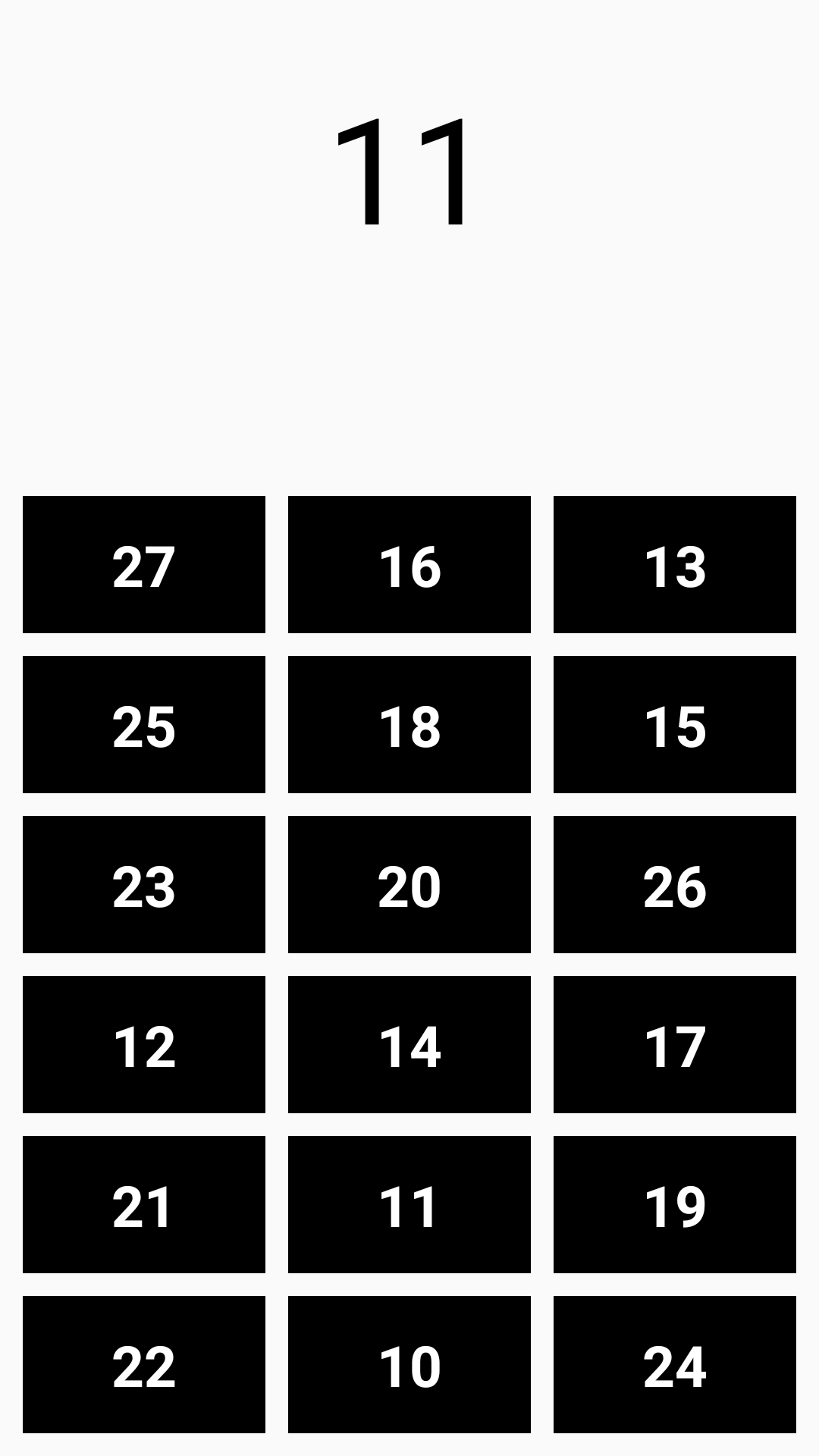}
	\caption{Screenshot of a test application for the pointing task}
	\label{fig:pointing_xp_screenshot}
\end{figure}

T1-T3 are everyday tasks users perform on a smartphone. They cover smartphone touch interaction of taps, swipes, and long press, on UI widgets such as keyboards, buttons and scrolling content. Users were asked to experience performing these tasks under six conditions, including under impairments (both using the physical hardware and the VR device). At the end of all three tasks (T1-T3), users completed the NASA-TLX\cite{nasatlx} survey to indicate their perceived workload during task performance. T4, on the other hand, is a controlled pointing task experiment. Participants were given a stimulus number and then asked to click on the button with the corresponding number, as \textit{quickly} and as \textit{precisely} as they could. (See Figure~\ref{fig:pointing_xp_screenshot} for a screenshot of the application used in this task.) Users repeated this task 80 times in succession, for each of the six conditions (A-F; see Table~\ref{tab:expt1taskblock}). We recorded the touch times and positions with the task app. We conducted a short semi-structured interview at the end of the study to understand users' experiences with, and perceptions of, the physical and virtual impairment simulations.

\textbf{Instruments}: We compared \name with a commercial physical impairment simulator \cite{lowvisionsimulator}.  To calibrate for \textit{visual acuity}, we adapted a test similar to a Snellen eye test chart \cite{snellen1873} - showing rows of letters with each lower row having a smaller font size. We first used the physical impairment simulator to obtain the minimum acceptable font size. Using the same test page in the VR, we applied the impairment and gradually adjusted the severity until we hit the minimum acceptable font size. To calibrate the inner circle of clarity for glaucoma, we implemented an app that allows us to adjust the diameter of a coloured circle. We then used the physical impairment simulator for glaucoma, and adjusted the coloured circle to the point in which the circle reaches the fringe for clarity. We then calibrated the virtual glaucoma simulation in a similar manner. Three independent measurements for visual acuity and circle of clarity were taken from the research team and averaged to determine the final calibration parameters of font size $=12~sp$ and diameter $=60~mm$.

\subsection{\name vs. Physical Smartphone}
We first investigate whether the VR-based interaction is a sufficiently faithful replica of the real-world interaction that a user would have with a regular smartphone, \emph{in the absence of any impairments}.

\textbf{Touch Accuracy:} In all six conditions, users were able to achieve high levels of button touch accuracy (see Table~\ref{tab:expt1accuracy}), with the accuracy being $98.8\%$ ($SD = 1.67$) when the users interacted unimpaired with the VR device. Comparing the accuracies between the physical smartphone and the VR device, we noted that the VR condition had an accuracy of $99.12\%$ ($SD = 1.32$) (across all 6 conditions), whereas the use of the physical smartphone provided $100\%$ accuracy. In terms of the location accuracy, we noted a difference of $2.28~mm$ ($SD = 2.98$) between the use of \name vs. a physical smartphone. This difference is well within the uncertainty associated with finger touch interactions, and thus demonstrates that user performance was equivalent across both \name and a physical smartphone.

\textbf{Perceived Workload:} NASA-TLX scores indicated that the users did perceive significant differences in their workload using \names, compared to use of the physical smartphone ($Z = 2.824$, $p = 0.005 < 0.05$). This does suggest that the navigating an app within the VR device does require greater cognitive effort than simply interacting with a regular smartphone. However, it is difficult to decipher whether this difference is due to \names-specific issues, or a general lack of familiarity with VR devices. 

We additionally investigated the subjective feedback captured by the semi-structured interview. $83\%$ (10) of the users reported perceiving increased latency while using \names, while 2 users indicated that they felt no noticeable latency difference. \textbf{However, all 12 users indicated that the performance of \name was ``acceptable", and they would be able to use the \name system for testing the usability of apps, as long as the apps do not require extremely low-latency interactions.} (3 users indicated that the system might not be usable for testing real-time games.)

\subsection{\name vs. Hardware Impairment Simulators}
We now study the performance of \name vis-a-vis impairments generated using commercially available hardware. Figure~\ref{fig:impairmentsimulationverhead} shows the overhead of \name under impairment conditions, demonstrating that \name is able to operate without significant performance loss even in the presence of impairments.

\begin{table}[]
	\centering
	\caption{Accuracy of Button Touch Across All Users}
	\label{tab:expt1accuracy}
	\begin{tabular}{|l|c|c|}
		\hline
		\rowcolor{gray!50}
		\textbf{Impairment}         & \textbf{Environment} & \textbf{Accuracy (SD) \%} \\ \hline
		& Physical             & 100                       \\ \cline{2-3} 
		\multirow{-2}{*}{None}      & Virtual              & 98.79 (1.67)              \\ \hline
		& Physical             & 100                       \\ \cline{2-3} 
		\multirow{-2}{*}{Cataracts} & Virtual              & 99.09 (1.36)              \\ \hline
		& Physical    & 100              \\ \cline{2-3} 
		\multirow{-2}{*}{Glaucoma}  & Virtual     & 99.49 (0.82)     \\ \hline
	\end{tabular}
\end{table}

\textbf{Touch Accuracy:}  Table~\ref{tab:expt1accuracy} enumerates the accuracy for the pointing task (T4) for two distinct impairments (\emph{Cataract} \& \emph{Glaucoma}), for both the VR-based \name system and the hardware impairment simulator. We see that, in the \emph{Cataract} condition, \name had a mean accuracy of $99.09\%$, which is virtually indistinguishable from that of the hardware device ($100\%$). A similar pattern was observed for the \emph{Glaucoma} impairment ($99.49$\% for \name vs. $100\%$ for Hardware). 
In terms of the location accuracy, we noted a difference of $1.7~mm$ ($SD = 1.9$) (for \emph{Cataract}) and $1.2~mm$ ($SD = 1.6$) (for \emph{Glaucoma}) between the use of \name vs. the impairment hardware. Once again, this difference is well within the uncertainty associated with finger touch interactions.  These results provide strong evidence that \name is able to emulate impairment conditions that are equivalent to that of dedicated, commercial hardware.

\textbf{Perceived Workload:} The numerical TLX scores indicated that there was no significant difference for \emph{Cataracts}; however, the difference for \emph{Glaucoma} was significant ($Z = 3.061$, $p = 0.002 < 0.05$), with users indicating a higher perceived workload for the VR device. 

\subsection{Motion sickness}
At the end of the user study, we asked each participant if they felt discomfort or unwell. Only two of the twelve participants reported slight motion sickness while using \names. Motion sickness may arise from: (1) the use of the VR display itself, and (2) the latency from \names. However, it is difficult to separate the two. 

The effects of motion sickness are notably minor in our current prototype of \names. The nature of our experimentation intensifies the use of the VR display, whereas practical use of \name is likely to be more interspersed between app redesigns. We further discuss how we may improve on latency in Section~\ref{ssection:dealingwithlatency} to reduce motion sickness that may result from the latency of \names.


\section{Related Work}
\textbf{Designing for Inclusiveness}. Newell et al.~\cite{newell2011user} pointed out that traditional user-centred design techniques provides little guidance for designing interfaces for elderly and disabled users due to the large variation amongst the type and degree of impairments. They also highlighted that the standard guidelines for designing disabled-friendly UIs are too general~\cite{newell1988human} and lacked empathy for users. For instance the WCAG 2.0 lists that the use of colour ``is not used as the only visual means of conveying information, indicating an action, prompting a response or distinguishing a visual element". This requires interpretation by the designer into specific designs in his application. Over the years, various accessibility design guidelines (such as WCAG 2.0~\cite{wcag20}, IBM Accessibility Checklist~\cite{ibmaccessibilitychecklist}, US Section 508 Standards~\cite{us508standards}) and tools (aChecker~\cite{gay2010achecker}) have been proposed and refined. However, the problems pointed out by Newell are remained unsolved to a large extent, which hinders elaborate design for a target user group with a specific impairment. 

\textbf{Simulated  Design}. There exists prior work on helping UI designers design better interfaces for people suffering from vision impairments. Higuchi et al. \cite{higuchi1999simulating} proposed a tool to simulate the visual capabilities of the elderly for the design of control panels, while Mankoff et al.~\cite{mankoff2005evaluating} developed a tool to simulate a user with visual and motor impairments on the desktop screen. SIMVIZ~\cite{ates2015immersive,werfel2016empathizing} uses the Oculus Rift VR device to simulate visual impairments to examine reading text on a smartphone. For audio modalities, Werfel et al.~\cite{werfel2016empathizing} simulated hearing ailments by using a pair of microphones with equalised headphones. 

Different from prior works, \name uses VR as the medium for immersive evaluation to 1) flexibly support wider groups of impaired users, and 2) allow naturalistic interactions with a \emph{mobile phone in a virtual environment}. This novel approach supports ecological validity in testing applications and is key for mobile apps which go beyond the static settings of previous work.

 While previous work has focused on simulation in single modality (visual or auditory), \name is able to flexibly combine modalities to support any application type, ailment (visual, auditory, motor) and usage environment. 



\textbf{System Support for Accessibility.} Modern mobile OSes provide accessibility support; in particular, it allows users with farsightedness to increase fonts and users with blindness to interact through vocal interfaces. Also, Zhong et al. enhanced Android accessibility for users with hand tremor by reducing fine pointing and steady tapping~\cite{zhong2015enhancing}. We believe \name will significantly expand basic accessibility support of commodity devices and accelerates the design and deployment of various accessibility add-ons for different impaired users. 

\textbf{Testing of Mobile Applications.} Recently there have been many systems, such as VanarSena~\cite{avindranath2014automatic}, AMC~\cite{Lee13}, Puma~\cite{hao2014puma}, DynoDroid~\cite{Machiry13}, DECAF~\cite{liu2014decaf}, AppsPlayground~\cite{rastogi2013appsplayground}, for automatically testing and identifying various types of UI and systems bugs in mobile applications.  \name takes a different approach in that we do not detect bugs after the application is developed and deployed.  Instead, we allow the designer to test early iterations of the designs rapidly.  In this way, we hope to reduce the pain of having to make significant UI changes at the end of the design cycle -- or worse, end with an application that cannot be used effectively by the target impaired demographic.



\section{Discussion}
Our current studies indicate the considerable promise of \names, as a mechanism for rapid and empathetic evaluation of app usability. We now discuss some additional studies and issues that we intend to explore further.

\subsection{User study with Designers}
We conducted a short user study with two mobile app developers to qualitatively examine \name in actual use. Both developers have previously worked to create an Android mobile application, which was used as the baseline for the study. The developers were tasked with redesigning the mobile app for the glaucoma-impaired under two conditions: 1) without \name, but with materials describing glaucoma and showing functionally accurate examples of glaucoma, and 2) with the same materials, and \name. Both developers agreed that \name helped them improve their designs over the baseline condition. The developers reported that \name allowed them to improve their designs in two ways: 1) they can focus their attention on re-designing particular problematic parts of the UI, and 2) they are able to appropriately calibrate their modifications (for instance increasing the font size may help, but text that is too large will also cause glaucoma sufferers to visually scan more, causing fatigue).

\subsection{Dealing with Latency Issues}\label{ssection:dealingwithlatency}
Our experimental studies indicate that users are able to utilise \name effectively for ``conventional" apps---i.e., those that typically involve sporadic interaction by users with UI elements, such as buttons and keyboards. 
The current end-to-end latency (of $\approx 200~msec$) is not an impediment for high-fidelity evaluation of such apps. 
However, the participants also indicated that this latency (lag between user actions and rendering in the VR display) would pose a problem for highly latency-sensitive applications, such as games. 
At present, it is thus appropriate to state that \name potentially needs additional optimisations to support such applications. 
The most obvious improvement would be to replace the default Android emulator with a faster, custom emulation engine--this is likely to reduce $\approx 100~msec$ of the delay budget.

The current implementation streams JPEG images (hand, emulator's screen) from the intermediary computer to the VR smartphone. We plan to adopt a low-latency video streaming codec such as H.265 HEVC~\cite{h265hevc}, which would help reduce networking and rendering latency. OS-level optimisations (e.g., preemptive priority for inter-component messages) may be needed to support even lower latency.

Recently, several works have proposed techniques for achieving high-quality VR experience on mobile devices~\cite{boos2016flashback, lai2017furion, abari2017movr}. \name could borrow some techniques to improve latency and video quality.

\subsection{User Performance with VR Devices}
Moreover, our user studies also indicated that the time for performing tasks (T1-T4) was marginally higher when using the VR environment, compared to the direct use of a real-world smartphone. More specifically, for the pointing task T4, there was an average difference of $654~msec$ in the task completion time using \names, compared to the smartphone. In addition, anecdotal comments suggest that continued use of the VR device, for longer-lived sessions, might pose additional usability challenges. For example, a couple of users indicated some minor muscle fatigue, most likely as a result of using a `heavy' VR device. It is an open question whether these issues will be mitigated over time, as VR devices become lighter and more ergonomic, and as users have greater familiarity with the use of VR devices.

\subsection{Advanced Uses of \name}
Our current implementation of \name supports the virtualisation of certain output modalities (specifically the display and audio) of the emulated app. The vision of \name can be extended to create other richer interaction modes, often blending virtual and augmented reality (AR) settings. As an example, certain emulation conditions may need to generate and integrate synthetic sensor traces, to replace the real sensor traces from the smartphone--e.g., to mimic the user's movement in locations, such as forests and mountains, the phone's real GPS trace would need to be replaced by a synthetic GPS trace as in \cite{min2015powerf, min2016pada}. Similarly, in some cases, the app itself might need to take \emph{inputs} from the VR world--e.g., if the app was being used to magnify certain objects embedded in the VR world. While such use cases can be supported, they will require enhancements to the current \name framework, and it is likely that the implementation may surface additional challenges, in terms of computational overhead and latency.

\subsection{Developing Impairment Filters and Profiles}
To demonstrate the viability of \names, we focused on demonstrating the ability to simulate visual impairments and in particular cataracts and glaucoma. As we explored, these impairments have functional aspects that are commonly employed to characterise them, such as visual acuity or contrast sensitivity, and are often accompanied by standard tests such as the Snellen eye test chart~\cite{snellen1873} and Pelli-Robson contrast sensitivity chart~\cite{pelli1988design} respectively. From examining the commercial physical impairment simulator and our experimentation, we believe that \name has the ability to functionally simulate other impairments.

We recognise two important directions that \name needs address to improve impairment simulation and use. First, impairment filters have to be developed in concert with medical professionals who are subject matter experts in the areas of the specific pathologies. This helps to develop a library of impairment filters. Second, with verified impairment filters, we may create \emph{impairment profiles}, which characterise groups of users with possibly overlapping requirements. For instance, a hypothetical impairment profile may calibrate for a demographic of a range of ages, sex, and percentage of the population who may have myopia and cataracts---both which affect visual acuity. With impairment profiles, app developers may easily select and understand the demographic to which they are designing for.

\section{Conclusion}
We presented the design and evaluation of \names, a framework that allows app developers to `step into the shoes' of impaired users, and perform an empathetic evaluation of their app interfaces. Our key idea is to utilise a virtual world (using a commodity VR device) to present an \emph{impaired} view of the app's interface, while allowing the user to interact naturally with a real commodity smartphone in the physical world. Overcoming the current computational limitations (of the VR device and the Android emulator) required us to make careful system choices, such as (i) appropriate tradeoffs between the resolution and frame rate for rendering the virtual smartphone, (ii) subsampling of the mesh representing the user's hand and (iii) scaling up the size of the virtual smartphone to overcome the lower resolution of the VR device. User studies show that \name is effective in (a) providing usability that is equivalent to using a real app (on a real smartphone), for applications that do not require ultra-low latency and (b) emulating impairments in a similar fashion to custom hardware devices. We believe that \name can be a powerful new paradigm for effective bidirectional integration between real-world user actions and virtual worlds, and that this can enable additional immersive applications beyond just `impairment emulation'.
\section{Acknowledgement}
We are thankful to our shepherd Prof. Xia Zhou and all anonymous reviewers for their valuable reviews.
This research is supported partially by Singapore Ministry of Education Academic Research Fund Tier 2 under research grant MOE2014-T2-1063, and by the National Research Foundation, Prime Minister's Office, Singapore under its IDM Futures Funding Initiative. All findings and recommendations are those of the authors and do not necessarily reflect the views of the granting agency, or SMU.

\bibliographystyle{ACM-Reference-Format}
\bibliography{mobisys18-empathd}

\end{document}